\documentclass{article}

\pdfoutput=1

\usepackage{amsfonts}
\usepackage{amsmath}
\usepackage{graphicx}

\begin{document}

\title{Grazing dynamics and dependence on initial conditions in certain
systems with impacts}
\author{Andrzej Okninski$^{1)}$ and Boguslaw Radziszewski$^{2)}$ \\
$^{1)}$Physics Division, $^{2)}$Institute of Processes Modeling\\
Politechnika Swietokrzyska, Al. 1000-lecia PP 7}
\maketitle

\begin{abstract}
Dynamics near the grazing manifold and basins of attraction for a motion of
a material point in a gravitational field, colliding with a moving
motion-limiting stop, are investigated. The Poincare map, describing
evolution from an impact to the next impact, is derived. Periodic points are
found and their stability is determined. The grazing manifold is computed
and dynamics is approximated in its vicinity. It is shown that on the
grazing manifold there are trapping as well as forbidden regions. Finally,
basins of attraction are studied.
\end{abstract}

Keywords: non-smooth dynamics; grazing; basins of attraction\bigskip

\section{\textbf{Introduction}}

Vibro-impacting systems are interesting examples of non-linear dynamical
systems, exhibiting new levels of complicated dynamics due to their
non-smoothness and have important industrial applications, see \cite%
{diBernardo2006} and references therein. Research in this field was
initiated by studies of impacting linear oscillators \cite{Fe1970} - \cite%
{Fo1994} as well as closely related accelerator models of particle physics
originating from the Fermi model \cite{Fermi1949} - \cite{Lu1996}. A very
characteristic feature of such systems is the presence of non-standard
bifurcations such as border-collisions \cite{Fe1970} - \cite{diBernardo1999}
and grazing impacts \cite{Pe1992}, \cite{Wh1987} - \cite{Fo1994}. Such
events often lead to complex chaotic motion, see \cite{diBernardo2006} for a
review of border-collision and grazing bifurcation phenomena. Little work
was done, however, on basins of attraction of different attractors typically
coexisting in impacting systems. Early results were obtained in \cite%
{Wh1987, Wh1992, No1992}. It seems however, that the structure of basins of
attraction, especially in the case of coexistence of grazing dynamics and
periodic orbits, deserves a further study.

In the present paper we investigate dynamics near the grazing manifold and
basins of attraction for a motion of a material point in a gravitational
field colliding with a moving motion-limiting stop. Typical example of such
dynamical system, related to the Fermi model, is a small ball bouncing
vertically on a vibrating table \cite{Ho1982} - \cite{Lu1996}. Since
evolution between impacts is expressed by a very simple formula the motion
in this system is easier to analyze than dynamics of impact oscillators.

The paper is organized as follows. In the second Section of this paper a
one-dimensional motion of a material point in a gravitational field,
colliding with a limiter, representing unilateral constraints, is
considered. It is assumed that, at impact, there is an inelastic rebound
with coefficient of restitution $0\leq R<1$ \cite{Goldsmith1960,
Pfeiffer1996}\emph{. }Next, the Poincar\'{e} section is defined and the
return map, describing evolution from an impact to the next impact, is
derived generalizing and extending our previous results \cite{Gajewski1987}.
In Section $3$ we find periodic points and determine their stability. In
Section $4$ the grazing manifold is computed. It turns out that the Poincar%
\'{e} map is not well defined on the grazing manifold. However, we
demonstrate that the map can be continued to the whole grazing manifold
after appropriate regularization procedure, except from some critical
points. Then the nonlinear equation for time of the next impact is
approximately solved and dynamics is linearized in the vicinity of the
grazing manifold. It is shown that on the grazing manifold there are
trapping as well as forbidden regions which are determined explicitly.
Finally, basins of attraction are studied in Section $5$. More exactly, we
investigate boundary between basins of grazing dynamics and periodic motion
which, for large velocities, becomes very complicated.

\section{Motion with impacts: the Poincar\'{e} section and return map\textbf{%
\ \ \label{B}}}

Let us consider motion of a material point moving vectically in a
gravitational field and colliding with a moving motion-limiting stop,
representing unilateral constraints. Let motion of the point in time
intervals between impacts be described by equation:%
\begin{equation}
m\ddot{x}=-mg,  \label{motion point}
\end{equation}%
where $\dot{x}\overset{df}{=}\frac{dx}{dt}$, while motion of the limiter is
expressed as:%
\begin{equation}
y=y\left( t\right) ,  \label{motion limiter}
\end{equation}%
with known function $y\left( t\right) $. We assume that $y$, $\dot{y}$, $%
\ddot{y}$ are continuous functions of $t$ for $t\in \lbrack 0,\infty )$.

Equations of impact are of form: 
\begin{subequations}
\label{IMPACT}
\begin{align}
x\left( \tau _{i}\right) & =y\left( \tau _{i}\right) ,  \label{impact1} \\
\dot{x}\left( \tau _{i}^{+}\right) -\dot{y}\left( \tau _{i}\right) & =-R%
\left[ \dot{x}\left( \tau _{i}^{-}\right) -\dot{y}\left( \tau _{i}\right) %
\right] ,  \label{impact2}
\end{align}%
where it is assumed that the limiter's mass is infinite. In (\ref{IMPACT}) $%
\tau _{i}$ denotes the instant of $i$-th impact, $\tau _{i}^{-}\left( \tau
_{i}^{+}\right) $ is the time just before (after) the $i$-th impact and $R$
is the coefficient of restitution, $0\leq R<1$ \cite{Goldsmith1960,
Pfeiffer1996}. The coefficient of restitution is a measure of the elasticity
of the collision between the material point and the limiter, and is also a
measure of energy dissipated during a collision. The coefficient of
restitution can be determined experimentally or computed on the basis of a
specific model of collisions. In the present work we shall assume that $R$
is known and constant.

It follows from the dynamics that $x(t)$ is a continuous function of time
while $\dot{x}\left( t\right) $, continuous within all time intervals $%
\left( \tau _{i-1}^{+},\ \tau _{i}^{-}\right) $, is subject to discontinuous
changes in time instants $\tau _{i}$\ described by the impact equation (\ref%
{impact2}). By $\dot{x}\left( \tau _{i}^{-}\right) $, $\dot{x}\left( \tau
_{i}^{+}\right) $ we understand the left-sided and right-sided limits of $%
\dot{x}\left( t\right) $ for $t\rightarrow \tau _{i}$, respectively, for
every $i=0,1,\ldots $ . In what follows we\ shall\ not\ assume \textit{a
priori} knowledge of $\tau _{i}$'s.

We shall now consider equation (\ref{motion point}) for $t\in \left( \tau
_{i}^{+},\ \tau _{i+1}^{-}\right) $. General solution reads: 
\end{subequations}
\begin{equation}
x\left( t\right) =-\tfrac{1}{2}gt^{2}+c_{1}^{\left( i\right)
}t+c_{2}^{\left( i\right) },  \label{solution2a}
\end{equation}

\begin{equation}
\dot{x}\left( t\right) =-gt+c_{1}^{\left( i\right) },  \label{solution2b}
\end{equation}
where $c_{1}^{\left( i\right) },$ $c_{2}^{\left( i\right) }$ are arbitrary
constants.

The constants $c_{1}^{\left( i\right) }$, $c_{2}^{\left( i\right) }$ in (\ref%
{solution2a}), (\ref{solution2b}) can be determined from the initial
conditions $x\left( \tau _{i}^{+}\right) $, $\dot{x}\left( \tau
_{i}^{+}\right) $:

\begin{equation}
c_{1}^{\left( i\right) }=\dot{x}\left( \tau _{i}^{+}\right) +g\tau
_{i},\quad c_{2}^{\left( i\right) }=\tfrac{1}{2}g\tau _{i}^{2}-c_{1}^{\left(
i\right) }\tau _{i}+x\left( \tau _{i}\right) .  \label{constants1}
\end{equation}

The solution (\ref{solution2a}), (\ref{solution2b}) can be now rewritten in
form: 
\begin{subequations}
\label{SOLUTION}
\begin{align}
x(t)& =-\tfrac{1}{2}gt^{2}+\left[ \dot{x}\left( \tau _{i}^{+}\right) +g\tau
_{i}\right] t-\tfrac{1}{2}g\tau _{i}^{2}-\dot{x}\left( \tau _{i}^{+}\right)
\tau _{i}+x(\tau _{i}),  \label{solution4a} \\
\dot{x}\left( t\right) & =-gt+\dot{x}\left( \tau _{i}^{+}\right) +g\tau _{i},
\label{solution4b}
\end{align}%
for each $i=0,1,...$and $t\in (\tau _{i}^{+},\ \tau _{i+1}^{-})$.
Substituting $t=\tau _{i+1}$ in (\ref{solution4a}) we get the formula for
the position of the material point after the $i+1$-th impact:

\end{subequations}
\begin{equation}
x\left( \tau _{i+1}\right) =x\left( \tau _{i}\right) -\tfrac{1}{2}g\left(
\tau _{i+1}-\tau _{i}\right) ^{2}+\dot{x}\left( \tau _{i}^{+}\right) \left(
\tau _{i+1}-\tau _{i}\right) .  \label{mapx}
\end{equation}

It is now possible, applying to (\ref{mapx}) the impact condition (\ref%
{impact1}), to obtain a nonlinear equation from which the time of the next
impact can be computed from the given impact time and the corresponding
velocity:%
\begin{equation}
y\left( \tau _{i+1}\right) =y\left( \tau _{i}\right) -\tfrac{1}{2}g\left(
\tau _{i+1}-\tau _{i}\right) ^{2}+\dot{x}\left( \tau _{i}^{+}\right) \left(
\tau _{i+1}-\tau _{i}\right) .  \label{mapta}
\end{equation}

On the other hand, substituting $t=\tau _{i+1}^{-}$ in (\ref{solution4b}) we
obtain the formula for the velocity of the material point just before $i+1$
-th impact:

\begin{equation}
\dot{x}\left( \tau _{i+1}^{-}\right) =\dot{x}\left( \tau _{i}^{+}\right)
-g\left( \tau _{i+1}-\tau _{i}\right) .  \label{velocity1}
\end{equation}
To obtain the velocity just after the $i+1$-th impact the equation (\ref%
{impact2}) is used:

\begin{equation}
\dot{x}\left( \tau _{i+1}^{+}\right) =-R\dot{x}\left( \tau _{i+1}^{-}\right)
+\left( 1+R\right) \dot{y}\left( \tau _{i+1}\right) .  \label{impact3}
\end{equation}%
Combining equations (\ref{impact3}) and (\ref{velocity1}) we obtain:%
\begin{equation}
\dot{x}\left( \tau _{i+1}^{+}\right) =R\left[ -\dot{x}\left( \tau
_{i}^{+}\right) +g\left( \tau _{i+1}-\tau _{i}\right) \right] +\left(
1+R\right) \dot{y}\left( \tau _{i+1}\right) .  \label{mapva}
\end{equation}

Finally, equations mapping the time of the $i$-th impact to the time of the $%
i+1$-th impact as well as the velocity of the material point just after the $%
i$-th impact to the velocity just after the $i+1$-th impact for arbitrary
motion of the limiter consist of Eqs.( \ref{mapta}), (\ref{mapva}): 
\begin{subequations}
\label{MAP}
\begin{align}
y\left( \tau _{i+1}\right) & =y\left( \tau _{i}\right) -\tfrac{1}{2}g\left(
\tau _{i+1}-\tau _{i}\right) ^{2}+\dot{x}\left( \tau _{i}^{+}\right) \left(
\tau _{i+1}-\tau _{i}\right) ,  \label{mapt1} \\
\dot{x}\left( \tau _{i+1}^{+}\right) & =-R\dot{x}\left( \tau _{i}^{+}\right)
+Rg\left( \tau _{i+1}-\tau _{i}\right) +\left( 1+R\right) \dot{y}\left( \tau
_{i+1}\right) .  \label{mapv1}
\end{align}%
With help of Eqs.(\ref{MAP}) and Eq.(\ref{mapx}) the solution (\ref{SOLUTION}%
) can be continued to the interval $\left( \tau _{i+1}^{+},\ \tau
_{i+2}^{-}\right) $.

Introducing non-dimensional variables: 
\end{subequations}
\begin{equation}
T_{i}=\omega \tau _{i},\ Y\left( T_{i}\right) =y\left( \tau _{i}\right) /a,\
V_{i}=\tfrac{2\omega }{g}\dot{x}\left( \tau _{i}^{+}\right) ,\ \lambda =%
\tfrac{2\omega ^{2}a}{g},  \label{units}
\end{equation}%
where $\omega $ and $a$ determine time and length scales, we obtain from (%
\ref{MAP}): 
\begin{subequations}
\label{MAPfinal}
\begin{align}
\lambda Y\left( T_{i+1}\right) & =\lambda Y\left( T_{i}\right) -\left(
T_{i+1}-T_{i}\right) ^{2}+V_{i}\left( T_{i+1}-T_{i}\right) ,  \label{mapt} \\
V_{i+1}& =-RV_{i}+2R\left( T_{i+1}-T_{i}\right) +\lambda \left( 1+R\right) 
\dot{Y}\left( T_{i+1}\right) ,  \label{mapv}
\end{align}%
where $\dot{Y}\overset{df}{=}\tfrac{dY}{dT}$. It follows that the map (\ref%
{MAPfinal}) depends on two control parameters only: $R$ and $\lambda $.

Assuming finally position of the limiter in form $y\left( t\right) =a\sin
\left( \omega t+\varphi \right) $, that is $Y\left( T\right) =\sin \left(
T+\varphi \right) $, we get: 
\end{subequations}
\begin{subequations}
\label{MAPTV}
\begin{align}
\lambda \sin \left( T_{i+1}+\varphi \right) & =\lambda \sin \left(
T_{i}+\varphi \right) -\left( T_{i+1}-T_{i}\right) ^{2}+V_{i}\left(
T_{i+1}-T_{i}\right) ,  \label{mapT} \\
V_{i+1}& =-RV_{i}+2R\left( T_{i+1}-T_{i}\right) +\lambda \left( 1+R\right)
\cos \left( T_{i+1}+\varphi \right) .  \label{mapVT}
\end{align}

The map (\ref{MAPTV}) is invariant under the translation $T_{i}\rightarrow
T_{i}+2\pi $ and thus the phase space $\left( T,\ V\right) $ is
topologically equivalent to the cylinder. Accordingly, Eqs. (\ref{MAPTV})
take a simpler form if a new variable $X_{i}\overset{df}{=}T_{i}+\varphi \ \
\left( mod \, 2\pi \right) $ is introduced: 
\end{subequations}
\begin{subequations}
\label{MAPXV}
\begin{align}
\lambda \sin \left( X_{i+1}\right) & =\lambda \sin \left( X_{i}\right)
-\left( X_{i+1}-X_{i}\right) ^{2}+V_{i}\left( X_{i+1}-X_{i}\right) ,
\label{mapX} \\
V_{i+1}& =-RV_{i}+2R\left( X_{i+1}-X_{i}\right) +\lambda \left( 1+R\right)
\cos \left( X_{i+1}\right) .  \label{mapV}
\end{align}

The map (\ref{MAPXV}) must meet several physical conditions to correspond
exactly to the original physical problem described by equations (\ref{motion
point}), (\ref{motion limiter}), (\ref{IMPACT}). First of all, we have to
choose the right solution out of infinitely many solutions of the nonlinear
equation (\ref{mapX}). We thus choose the solution $X_{i+1}$ with the
smallest nonnegative difference $X_{i+1}-X_{i}$. Moreover, since $V_{i}$ is
the velocity of the material point just after the impact, it must not be
smaller than the velocity of the limiter. Since the quantity 
\end{subequations}
\begin{equation}
V_{i}-\lambda \cos \left( X_{i}\right) =\tfrac{2\omega }{g}\left( \dot{x}%
\left( \tau _{i}^{+}\right) -\dot{y}\left( \tau _{i}\right) \right) ,
\label{defu}
\end{equation}%
is the relative velocity of the material point in non-dimensional units we
implement equations (\ref{MAPXV}) with the following physical conditions: 
\begin{subequations}
\label{CONDITIONS}
\begin{eqnarray}
&&%
\begin{array}{cc}
\delta _{i}\overset{df}{=} & X_{i+1}-X_{i}\geq 0%
\end{array}%
,  \label{condition1} \\
&&%
\begin{array}{cc}
w_{i}\overset{df}{=} & V_{i}-\lambda \cos \left( X_{i}\right) \geq 0%
\end{array}%
,  \label{condition2}
\end{eqnarray}%
which define the physical phase space $\left( X,\ V\right) $.

Equations similar to (\ref{MAP}) were derived in \cite{Lu1996} for position
of the limiter in the special form $y\left( \tau \right) =a\sin \left(
\omega \tau +\varphi \right) $ and correct stability conditions for fixed
points of the map were determined. However, the authors did not use the
non-dimensional variables as in (\ref{MAPfinal}) or (\ref{MAPXV}) and hence
their map contained redundant control parameters. Moreover, they did not
compute the map for the velocity just after the impact, $\dot{x}\left( \tau
_{i}^{+}\right) $, for which the important inequality $\dot{x}\left( \tau
_{i}^{+}\right) \geq \dot{y}\left( \tau _{i}\right) $ or (\ref{condition2})
holds, but for the velocity just before the impact, $\dot{x}\left( \tau
_{i}^{-}\right) $.

Computations of impact times $T_{i}$'s from nonlinear equation (\ref{mapT})
are complicated. The problem simplifies greatly for motions for which the
approximate equality $x\left( T_{i+1}\right) \cong x\left( T_{i}\right) $
or, equivalently, $y\left( T_{i+1}\right) \cong y\left( T_{i}\right) $\
holds. The condition $y\left( T_{i+1}\right) =y\left( T_{i}\right) $ leads
to $\lambda \sin \left( T_{i+1}+\varphi \right) =\lambda \sin \left(
T_{i}+\varphi \right) $ which applied to (\ref{mapT}) yields $%
T_{i+1}-T_{i}=V_{i}$. Hence Eqs.(\ref{MAPTV}) reduce to the following
approximate map: 
\end{subequations}
\begin{subequations}
\label{MAP(Holmes)}
\begin{align}
T_{i+1}& =T_{i}+V_{i},  \label{map1b} \\
V_{i+1}& =RV_{i}+\lambda \left( 1+R\right) \cos \left( T_{i}+V_{i}+\varphi
\right) ,  \label{map2b}
\end{align}%
equivalent for $\varphi =\pi $ to the map introduced by Holmes \cite{Ho1982}
(note that $\lambda \left( 1+R\right) $ is the control parameter $\gamma $
of Holmes). The map (\ref{MAP(Holmes)}) was studied theoretically \cite%
{Ho1982} while experimental results were reported in \cite{Pi1986, Ko1988}.

\section{Fixed points and their stability}

Since the map (\ref{MAPXV}) is $2\pi $ periodic in $X$ the fixed points ($1$
- cycles) are found by substituting $X_{i+1}^{\left( n\right)
}-X_{i}^{\left( n\right) }=2n\pi $. Accordingly, we get: 
\end{subequations}
\begin{subequations}
\label{FIXED}
\begin{align}
V_{\ast }^{\left( n\right) }& =2\pi n,  \label{fixedV} \\
\cos \left( X_{\ast }^{\left( n\right) }\right) & =\tfrac{2\pi n(1-R)}{%
\lambda \left( 1+R\right) },  \label{fixedX}
\end{align}%
where $\left( X_{\ast }^{\left( n\right) },\ V_{\ast }^{\left( n\right)
}\right) $ is a fixed point on the cylinder $S\times \mathbb{R}$. In (\ref%
{FIXED}) $n$ is an arbitrary non-zero integer such that: 
\end{subequations}
\begin{equation}
-1\leq \tfrac{2\pi n(1-R)}{\lambda \left( 1+R\right) }\leq 1\qquad \left(
n=\pm 1,\ \pm 2,\ \ldots \right) .  \label{inequality}
\end{equation}

The periodic motion of the material point is thus possible when the
conditions (\ref{FIXED}), (\ref{inequality}) are fulfilled.

From (\ref{fixedV}) and (\ref{fixedX})\ we have, for a given $n$, the pair
of fixed points:%
\begin{equation}
V_{\ast }^{\left( n\right) }=2\pi n,\quad X_{\ast 1}^{\left( n\right)
}=\arccos \left( \tfrac{2\pi n(1-R)}{\lambda \left( 1+R\right) }\right) ,\
X_{\ast 2}^{\left( n\right) }=2\pi -X_{\ast 1}^{\left( n\right) }.
\label{solfixed}
\end{equation}

Let us notice here that the fixed points of the map (\ref{MAPfinal}) are
exactly the same as the fixed points of the approximate map (\ref%
{MAP(Holmes)}) but, of course, their stabilities should differ.

Stability of periodic solutions (\ref{solfixed}) can be investigated by
substituting in (\ref{MAPXV}) $X_{i}=X_{\ast p}^{\left( n\right) }+d_{i}$, $%
V_{i}=V_{\ast }^{\left( n\right) }+u_{i}$ with small deviations $d_{i}$, $%
u_{i}$ from the fixed point. In all formulae below we have $p=1,2$. After
some rearrangements we get equations for the perturbations $d_{i}$, $u_{i}$: 
\begin{subequations}
\label{DU1}
\begin{align}
\hspace{-0.18in}\lambda \sin \left( d_{i+1}+X_{\ast p}^{\left( n\right)
}\right) & =\lambda \sin \left( d_{i}+X_{\ast p}^{\left( n\right) }\right)
-\left( d_{i+1}-d_{i}\right) ^{2}-4\pi n\left( d_{i+1}-d_{i}\right) +
\label{d1} \\
& \hspace{0.14in}+2\pi n\left( d_{i+1}-d_{i}\right) +2\pi nu_{i},\medskip 
\notag \\
u_{i+1}& =-Ru_{i}+2R\left( d_{i+1}-d_{i}\right) +  \label{u1} \\
& \hspace{0.14in}+\lambda \left( 1+R\right) \left[ \cos \left(
d_{i+1}+X_{\ast p}^{\left( n\right) }\right) -\cos \left( X_{\ast p}^{\left(
n\right) }\right) \right] .  \notag
\end{align}

The system of equations (\ref{DU1}) has the trivial solution $d_{i}=u_{i}=0$
. Linearizing Eqs.(\ref{DU1}) around the trivial solution we obtain: 
\end{subequations}
\begin{eqnarray}
\hspace{-0.3in}\lambda d_{i+1}\cos \left( X_{\ast p}^{\left( n\right)
}\right) &=&\lambda d_{i}\cos \left( X_{\ast p}^{\left( n\right) }\right)
-2\pi n(d_{i+1}-d_{i})+2\pi nu_{i},  \label{d2} \\
u_{i+1} &=&-Ru_{i}+2R(d_{i+1}-d_{i})-\lambda (1+R)d_{i+1}\sin \left( X_{\ast
p}^{\left( n\right) }\right) ,  \label{u2}
\end{eqnarray}%
and substituting $X_{\ast p}^{\left( n\right) }$ from Eq.(\ref{solfixed}) we
obtain the linearized map defined via the stability (Jacobi) matrix $\mathbb{%
S}$: 
\begin{subequations}
\label{D}
\begin{align}
& 
\begin{array}{c}
\left( 
\begin{array}{c}
d_{i+1} \\ 
u_{i+1}%
\end{array}%
\right) =\mathbb{S}\left( 
\begin{array}{c}
d_{i} \\ 
u_{i}%
\end{array}%
\right)%
\end{array}%
,\bigskip  \label{d3} \\
& \hspace{0.18in}%
\begin{array}{c}
\mathbb{S}=\left( 
\begin{array}{cc}
1 & \tfrac{1+R}{2}\medskip \\ 
-\mu _{p}^{\left( n\right) } & R^{2}-\tfrac{1+R}{2}\mu _{p}^{\left( n\right)
}%
\end{array}%
\right)%
\end{array}%
,  \label{d4}
\end{align}%
where $\mu _{1,2}^{(n)}=\pm \sqrt{\lambda ^{2}\left( 1+R\right) ^{2}-4\pi
^{2}n^{2}\left( 1-R\right) ^{2}}$. Let us note that due to (\ref{inequality}%
) $\mu _{1,2}^{(n)}$ are real.

The trivial solution $d_{i}=u_{i}=0$ is asymptotically stable if and only if
the roots $\Lambda _{1}$, $\Lambda _{2}$ of the characteristic equation of
the stability matrix: 
\end{subequations}
\begin{subequations}
\label{ChE}
\begin{align}
& 
\begin{array}{cc}
\det \left( 
\begin{array}{cc}
1-\Lambda & \frac{1+R}{2}\medskip \\ 
-\mu _{p}^{\left( n\right) } & \left( R^{2}-\tfrac{1+R}{2}\mu _{p}^{\left(
n\right) }\right) -\Lambda%
\end{array}%
\right) & =\Lambda ^{2}+\alpha _{p}\Lambda +\beta =0,%
\end{array}%
\medskip  \label{CharEquation} \\
& 
\begin{array}{cc}
\alpha _{p}=\tfrac{1+R}{2}\mu _{p}^{\left( n\right) }-1-R^{2},\ \beta =R^{2},
& 
\end{array}
\label{def}
\end{align}%
fulfill inequalities $\left\vert \Lambda _{1,2}\right\vert <1$. Using the
Schur-Cohn criterion \cite{Mickens1990} we get: 
\end{subequations}
\begin{equation}
\beta <1,\ \left\vert \alpha _{p}\right\vert <\beta +1.  \label{inequalities}
\end{equation}

Let us first consider the first fixed point $\left( X_{\ast 1}^{\left(
n\right) },\ V_{\ast }^{\left( n\right) }\right) \equiv P_{s}^{\left(
n\right) }$. For $p=1$ we obtain from Eqs.(\ref{ChE}), (\ref{inequalities})
conditions guaranteeing asymptotic stability of $P_{s}^{\left( n\right) }$: 
\begin{subequations}
\label{ST}
\begin{align}
& 
\begin{array}{c}
R<1%
\end{array}%
,\smallskip  \label{s1} \\
& 
\begin{array}{c}
\lambda _{cr_{1}}^{\left( n\right) }\equiv \tfrac{2\pi n(1-R)}{1+R}\leq
\lambda \leq \lambda _{cr_{2}}^{\left( n\right) }\equiv \tfrac{2}{(1+R)^{2}}%
\sqrt{n^{2}\pi ^{2}(1-R^{2})^{2}+4(1+R^{2})^{2}}%
\end{array}%
,  \label{s2}
\end{align}%
assuming that the inequality (\ref{inequality}) is fulfilled. It turns out
that stability of the fixed points of the exact map (\ref{MAPfinal}) is
indeed different than the stability of these points for the approximate map (%
\ref{MAP(Holmes)}), see \cite{Ho1982}.

On the other hand, the second fixed point $\left( X_{\ast 2}^{\left(
n\right) },\ V_{\ast }^{\left( n\right) }\right) \equiv P_{u}^{\left(
n\right) }$ is unstable. Indeed, for $\lambda \geq \lambda _{cr_{1}}^{\left(
n\right) }\ $and $0\leq R<1$ the discriminant of the characteristic equation
(\ref{ChE}) is positive so that $\Lambda _{1},\ \Lambda _{2}\ $are real.
Furthermore, the inequalities $0\leq \Lambda _{1}\Lambda _{2}=R^{2},\
\Lambda _{1}+\Lambda _{2}>1+R^{2}$ result from (\ref{ChE}) for $p=2$ and $%
\lambda >\lambda _{cr_{1}}^{\left( n\right) }$. It follows finally, that $%
0\leq \Lambda _{1}\leq R^{2}<1$ and$\ 1<\Lambda _{2}$. The fixed point $%
P_{u}^{\left( n\right) }$ is thus a saddle. This result and the inequality (%
\ref{s2}) were also derived in \cite{Lu1996}.

\section{Grazing dynamics}

\subsection{Bouncing ball}

The map (\ref{MAPXV}) has also the manifold of fixed points, defined by the
condition: 
\end{subequations}
\begin{equation}
X_{i+1}=X_{i}\equiv X_{\ast },\ V_{i+1}=V_{i}=V_{\ast }=\lambda \cos X_{\ast
},  \label{manifold}
\end{equation}%
what is verified by the direct substitution into Eq.(\ref{MAPXV}). We shall
refer to the manifold (\ref{manifold}) as the grazing manifold. In this
Section we shall analyze dynamics near the grazing manifold.

Let us first notice that a function $X_{i+1}=f\left( X_{i},\ V_{i}\right) $
is defined implicitly by Eq.(\ref{mapX}):%
\begin{equation}
\begin{array}{lll}
F\left( X_{i+1},\ X_{i},\ V_{i}\right) & \overset{df}{=} & \lambda \sin
X_{i+1}-\lambda \sin X_{i}+\left( X_{i+1}-X_{i}\right) ^{2}+\smallskip \\ 
&  & -\left( X_{i+1}-X_{i}\right) V_{i}=0.%
\end{array}
\label{implicit1}
\end{equation}

Therefore since%
\begin{equation}
\frac{\partial F\left( X_{i+1},\ X_{i},\ V_{i}\right) }{\partial X_{i+1}}%
=\lambda \cos X_{i+1}+2\left( X_{i+1}-X_{i}\right) -V_{i},  \label{implicit2}
\end{equation}%
then on the grazing manifold $\partial F\left( X_{i+1},\ X_{i},\
V_{i}\right) /\partial X_{i+1}=0$, see Eq.(\ref{manifold}), and thus $%
X_{i+1}=f\left( X_{i},\ V_{i}\right) $\ cannot be defined therein as follows
from the implicit function theorem \cite{Spivak1965}.

The map can be however continued on the whole grazing manifold except from
some critical points. Let us first notice that since there was impact at
time $X_{i}$, it follows that $X_{i+1}=X_{i}$ is a trivial solution of
equation (\ref{implicit1}), corresponding to the imposed initial condition.
This trivial solution is the source of singular behaviour of the map on the
grazing manifold. We can get rid of the trivial solution dividing the
function $F\left( X_{i+1},\ X_{i},\ V_{i}\right) $ by $X_{i+1}-X_{i}$ and
expressing as a function of $V_{i}$, $X_{i}$ and $\delta _{i}\equiv
X_{i+1}-X_{i}$:

\begin{subequations}
\label{implicit3}
\begin{align}
& 
\begin{array}{c}
G\left( \delta _{i},\ X_{i},\ V_{i}\right) \overset{df}{=}\frac{F\left(
X_{i+1},\ X_{i},\ V_{i}\right) }{X_{i+1}-X_{i}},%
\end{array}%
\smallskip  \label{implicit3a} \\
& 
\begin{array}{c}
G\left( \delta _{i},\ X_{i},\ V_{i}\right) =\tfrac{\lambda }{\delta _{i}}%
\left( \sin \delta _{i}\cos X_{i}+\left( \cos \delta _{i}-1\right) \sin
X_{i}\right) +\delta _{i}-V_{i}=0.%
\end{array}
\label{implicit3b}
\end{align}

Let us notice that the function $G\left( \delta _{i},\ X_{i},\ V_{i}\right) $
is well defined, the limiting case $\delta _{i}\rightarrow 0$ including.
Equation (\ref{implicit3}) defines implicitly a function $\delta
_{i}=g\left( X_{i},\ V_{i}\right) $. Since 
\end{subequations}
\begin{equation}
\frac{\partial G\left( \delta _{i},\ X_{i},\ V_{i}\right) }{\partial \delta
_{i}}=\left( -\tfrac{1}{2}\lambda \sin X_{i}+1\right) +O\left( \delta
_{i}\right) ,  \label{implicit4}
\end{equation}%
it follows that $\delta _{i}=g\left( X_{i},\ V_{i}\right) $ can be defined
everywhere on the grazing manifold, i.e. for $\delta _{i}\rightarrow 0$,
except from critical points:%
\begin{equation}
1-\tfrac{1}{2}\lambda \sin X_{cr}=0,\   \label{critical}
\end{equation}%
which appear for $\lambda \geq 2$, where the assumption of the implicit
function theorem, $\partial G\left( \delta _{i},\ X_{i},\ V_{i}\right)
/\partial \delta _{i}\neq 0$, is not fulfilled. It follows easily from Eqs.( %
\ref{manifold}), (\ref{critical}) that in critical points $V_{cr}\left(
\lambda \right) =\pm \sqrt{\lambda ^{2}-4}$.

It is important that while the quantity $\delta _{i}$ is non-negative, see (%
\ref{condition1}), it is also small near the grazing manifold:%
\begin{equation}
0\leq \delta _{i}\equiv X_{i+1}-X_{i}<1.  \label{assumption1}
\end{equation}%
It is thus natural to expand $G\left( \delta _{i},\ X_{i},\ V_{i}\right) $
in a power series of $\delta _{i}$:%
\begin{equation}
\begin{array}{lll}
G\left( \delta _{i},\ X_{i},\ V_{i}\right) & \hspace{-0.1in}= & \hspace{%
-0.1in}\left( \lambda \cos X_{i}-V_{i}\right) +\left( 1-\tfrac{1}{2}\lambda
\sin X_{i}\right) \delta _{i}-\tfrac{1}{6}\lambda \cos X_{i}\delta
_{i}^{2}+\smallskip \\ 
&  & +\frac{1}{24}\lambda \sin X_{i}\delta _{i}^{3}+\ldots \ =0,%
\end{array}
\label{approx}
\end{equation}%
which converges under assumption (\ref{assumption1}). Very close to the
grazing manifold $\delta _{i}\ll 1$ and we can neglect higher-order terms in
(\ref{approx}) to get approximate formula for the function $g\left( X_{i},\
V_{i}\right) $:%
\begin{equation}
\delta _{i}=g\left( X_{i},\ V_{i}\right) =\tfrac{V_{i}-\lambda \cos X_{i}}{1-%
\tfrac{1}{2}\lambda \sin X_{i}}+O\left( \delta _{i}^{2}\right) .
\label{maptA}
\end{equation}

Eq.(\ref{maptA}) in the initial variables (\ref{units}) has the following
form:%
\begin{equation}
\tau _{i+1}-\tau _{i}=\dfrac{-2v_{i}}{A_{i}},\qquad v_{i}\overset{df}{=}\dot{%
x}\left( \tau _{i}^{+}\right) -\dot{y}\left( \tau _{i}\right) ,\ A_{i}%
\overset{df}{=}\ddot{x}\left( \tau _{i}^{+}\right) -\ddot{y}\left( \tau
_{i}\right) ,  \label{maptB}
\end{equation}%
where $v_{i}$ and $A_{i}$ are relative post-impact velocity and
acceleration, respectively. We shall demonstrate in the next Subsection that
the formula (\ref{maptB}) is more general.

The approximation (\ref{maptA}) breaks down in critical points on the
grazing manifold defined by (\ref{critical}), that is $v_{i}=0,$ $A_{i}=0$.
On the other hand, we can use (\ref{maptA}) near the grazing manifold
wherever the consistency conditions%
\begin{equation}
0<\delta _{i}=\tfrac{V_{i}-\lambda \cos X_{i}}{1-\tfrac{1}{2}\lambda \sin
X_{i}}<1,  \label{cond}
\end{equation}%
are fulfilled, see (\ref{assumption1}). It follows from inequality (\ref%
{condition2}) that the condition (\ref{cond}) is violated for%
\begin{equation}
1-\tfrac{1}{2}\lambda \sin X_{i}<0.  \label{forbidden}
\end{equation}%
Therefore the region (\ref{forbidden}) of the grazing manifold is
non-physical and thus forbidden, see Fig. $1$. On the other hand, Eq.(\ref%
{maptB}) shows that in the forbidden region $v_{i}=0$ and $A_{i}>0$, c.f.
the definitions (\ref{maptB}), and thus this region is repelling. The
acceleration function was also used to analyze motion of impact oscillator 
\cite{Wh1987, Wh1992, No1991}.

The map (\ref{MAPXV}) can be simplified now in the neighbourhood of the
grazing manifold. Equation (\ref{mapX}) can be approximated by (\ref{maptA})
and we rewrite (\ref{MAPXV}) in the following form: 
\begin{subequations}
\label{GRAZING1}
\begin{align}
w_{i}& =\left( 1-\frac{1}{2}\lambda \sin X_{i}\right) \delta _{i}\
,\smallskip  \label{grazing1a} \\
w_{i+1}& =-Rw_{i}+2R\delta _{i}+\lambda R\left( \cos X_{i+1}-\cos
X_{i}\right) ,  \label{grazing1b}
\end{align}%
where $w_{i}\equiv V_{i}-\lambda \cos X_{i}$, $\delta _{i}\equiv
X_{i+1}-X_{i}$. Expanding now $\cos \left( X_{i+1}\right) =\cos \left(
X_{i}+\delta _{i}\right) $ and neglecting higher-order terms in $\delta _{i}$
we obtain the map: 
\end{subequations}
\begin{subequations}
\label{GRAZING2}
\begin{align}
w_{i}& =\left( 1-\frac{1}{2}\lambda \sin X_{i}\right) \delta _{i}\
,\smallskip  \label{grazing2a} \\
w_{i+1}& =-Rw_{i}+2R\left( 1-\tfrac{1}{2}\lambda \sin X_{i}\right) \delta
_{i},  \label{grazing2b}
\end{align}%
which, finally, takes a very simple form: 
\end{subequations}
\begin{equation}
w_{i+1}=Rw_{i}.  \label{grazing3}
\end{equation}%
Hence for $R<1$\ this part of the grazing manifold where the condition (\ref%
{cond}) is fulfilled\ is attracting. Therefore $w_{i}\rightarrow 0$ and it
follows from (\ref{grazing2a}) that also $\delta _{i}\rightarrow 0$ for $1-%
\frac{1}{2}\lambda \sin X_{i}\rightarrow 1-\frac{1}{2}\lambda \sin X_{\ast
}>0$, i.e outside critical points. In the final stage of grazing collisions
are thus so rapid that the motion of the limiter can be neglected.

We shall refer to the region on the grazing manifold $1-\tfrac{1}{2}\lambda
\sin X_{i}<0\ $as the forbidden or repelling region while the attracting
part of the grazing manifold will be referred to as the trapping region, see
Fig. $1$.\medskip

\begin{figure}[th!]
\center \includegraphics[width=10cm, height=8cm]{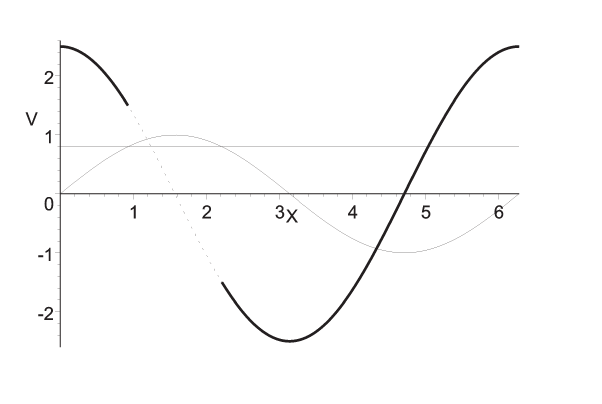}
\caption{Trapping (medium solid line) and forbidden (dotted line) regions of the
grazing manifold $V=\lambda \cos X$; $V=\frac{2}{\lambda }$ and $V=\sin X$
(thin solid lines),$\ \lambda =2.5$. Critical points are at the ends of
dotted lines.}
\label{F1}
\end{figure}

\subsection{Impact oscillator}

In the present Subsection we shall work in a general setting, proposed by
Nordmark \cite{No1991}\emph{. }The equation of motion is:%
\begin{equation}
\frac{d^{2}x}{dt^{2}}=A\left( x,\frac{dx}{dt},t\right) ,  \label{Nord1}
\end{equation}%
with periodic acceleration function:%
\begin{equation}
A\left( x,\frac{dx}{dt},t+2\pi \right) =A\left( x,\frac{dx}{dt},t\right) .
\label{Nord2}
\end{equation}%
The motion of the material point is confined to $x\geq x_{c}$ and the
stationary impact boundary is given by $x=x_{c}$. The impact law is of form $%
v_{i+1}=G\left( v_{i}\right) $, $G\left( 0\right) =0$, see for example Eq.( %
\ref{impact2}). This equation of motion can describe linear impact
oscillators as well as a ball bouncing on a vibrating table, see Section $%
5.2.3$ of \cite{Lamba1995} for the transformation relating systems with a
moving boundary to systems with a stationary stop.

The solution of (\ref{Nord1}) can be written in form:%
\begin{equation}
x=x\left( t,x_{c},v_{0},t_{0}\right) ,  \label{solution}
\end{equation}%
with initial conditions $x\left( t_{0},x_{c},v_{0},t_{0}\right) =x_{c}$, $%
\left. \frac{dx\left( t,x_{c},v_{0},t_{0}\right) }{dt}\right\vert
_{t=t_{0}^{+}}=v_{0}\geq 0$, i.e. we assume that the impact occurred at $%
t=t_{0}$. Let us also assume that the next impact occurs at $t=t_{1}$:%
\begin{equation}
F\left( t_{1},x_{c},v_{0},t_{0}\right) \overset{df}{=}x\left(
t_{1},x_{c},v_{0},t_{0}\right) -x_{c}=0,  \label{impact}
\end{equation}%
where the time of the second impact is defined by (\ref{impact}) as an
implicit function $t_{1}=f\left( x_{c},v_{0},t_{0}\right) $. Obviously, due
to the initial condition, it follows that $t_{1}=t_{0}$ is the trivial
solution of (\ref{impact}) and can be thus ignored. We shall next assume
that $t_{1}-t_{0}$ is small - this assumption is clearly fulfilled in the
phase of grazing. Expanding the function $F$ we get:%
\begin{equation}
x_{c}+\left( t_{1}-t_{0}\right) \left. \frac{dx}{dt_{1}}\right\vert
_{t=t_{0}^{+}}+\tfrac{1}{2}\left( t_{1}-t_{0}\right) ^{2}\left. \frac{d^{2}x%
}{dt_{1}^{2}}\right\vert _{t=t_{0}^{+}}+O\left( \left( t_{1}-t_{0}\right)
^{3}\right) -x_{c}=0.  \label{expansion}
\end{equation}%
Neglecting the trivial solution $t_{1}=t_{0}$ we obtain from (\ref{expansion}
) the approximate formula for the time interval elapsing between two
subsequent impacts:%
\begin{equation}
t_{1}-t_{0}=\frac{-2v_{0}}{A_{0}},  \label{time}
\end{equation}%
where $A_{0}\equiv \left. \frac{d^{2}x}{dt_{1}^{2}}\right\vert
_{t=t_{0}^{+}}=A\left( x_{c},v_{0},t_{0}^{+}\right) $.

\section{Basins of attraction}

We have studied dependence of the dynamics on the control parameters. The
bifurcation diagram is shown in Fig. $2$ for $R=0.85$ and $0\leq \lambda
\leq 2.5$. It can be seen that for $\lambda <\lambda _{cr_{1}}^{\left(
1\right) }=0.5094474\ldots \ $only the grazing dynamics is possible. For $%
\lambda >\lambda _{cr_{1}}^{\left( 1\right) }$ the first fixed point $%
P_{s}^{\left( 1\right) }$ exists and is attractive and the second fixed
point $P_{s}^{\left( 2\right) }$ exists and is attractive for $\lambda
>\lambda _{cr_{1}}^{\left( 2\right) }=1.0188949\ldots \ $. The first of $3$
- cycles, which typically accompany $1$ - cycles, appears for $\lambda \cong
1.78$ while several small attractors can be seen for $\lambda >1.5$.\medskip

\begin{figure}[th!]
\center \includegraphics[width=10cm, height=8cm]{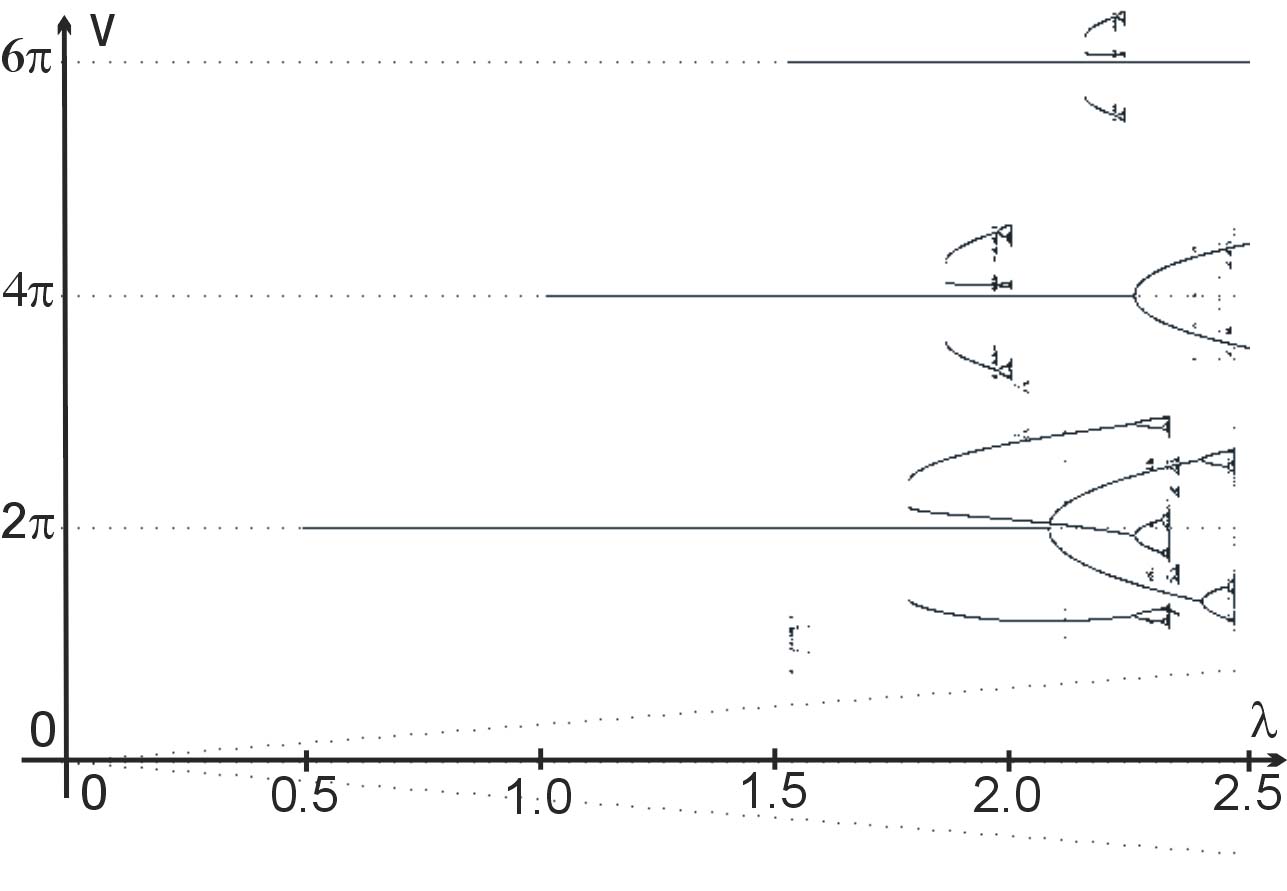}
\caption{Bifurcation diagram, $R=0.85,\ 0\leq \lambda \leq 2.5$. Dotted lines enclose
triangular region of grazing dynamics with $-\lambda \leq V\leq \lambda $.}
\label{F2}
\end{figure}

It follows that for $\lambda _{cr_{1}}^{\left( 1\right) }<\lambda <\lambda
_{cr_{1}}^{\left( 2\right) }$ there are two coexisting attractors only: the
grazing manifold and the first fixed point $P_{s}^{\left( 1\right) }$
.\medskip

\begin{figure}[th!]
\center \includegraphics[width=10cm, height=8cm]{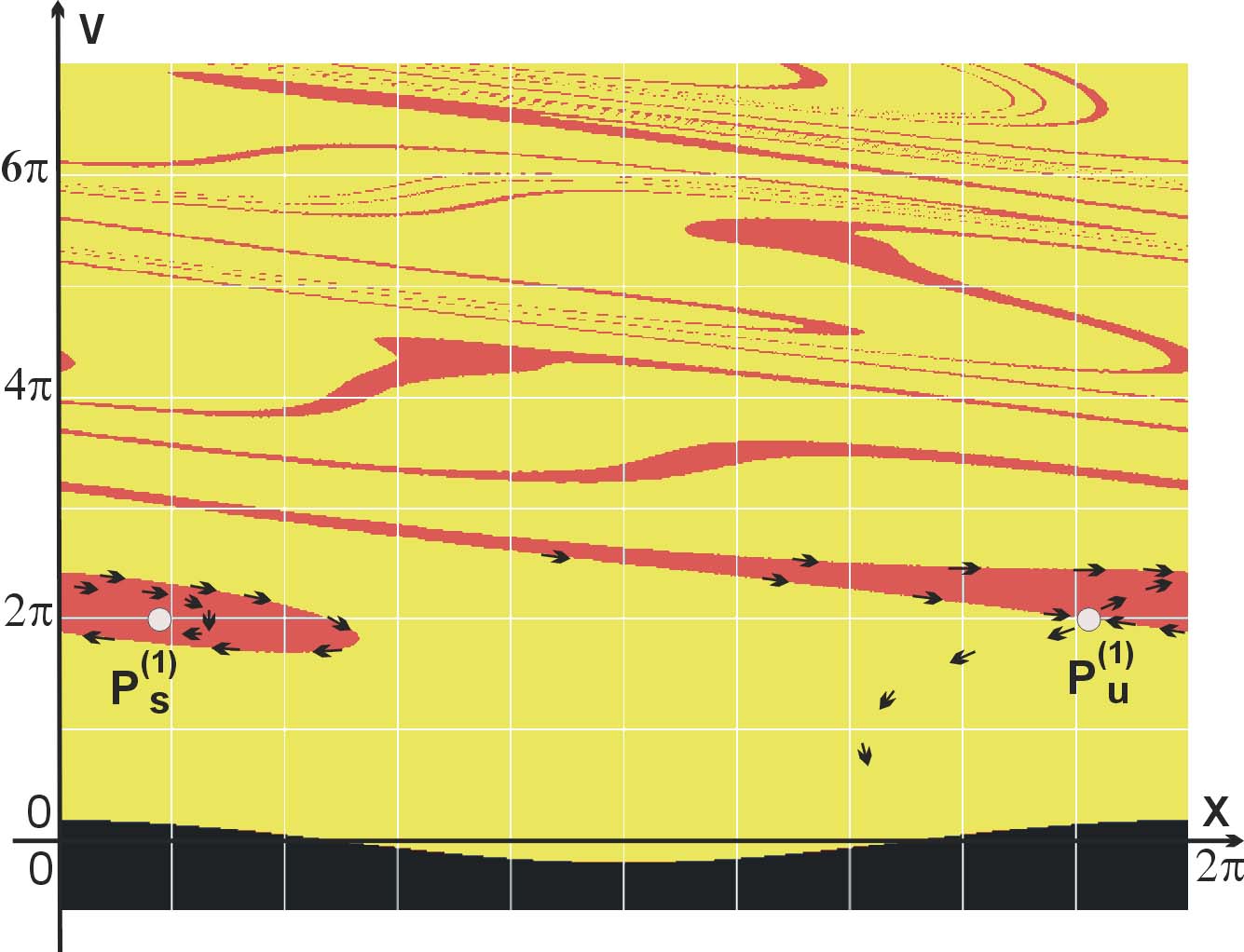}
\caption{Basins of attraction, $R=0.85,\ \lambda =0.6$.}
\label{F3}
\end{figure}

Basins of attraction of these two attractors are shown in Fig.$\,3$ above
for $R=0.85,\ \lambda =0.6$. In the Figures $3-5$ basins of grazing dynamics
are yellow, basins of $P_{s}^{\left( 1\right) }$ are coloured red and basin
of $P_{s}^{\left( 2\right) }$ is green. Arrows denote direction of motion on
stable and unstable manifolds of unstable fixed points $P_{u}^{\left(
1\right) }$, $P_{u}^{\left( 2\right) }$. Boundaries of the attractor's
basins are determined by the nature of the unstable fixed point $%
P_{u}^{\left( 1\right) }$. This unstable fixed point appears for growing $%
\lambda $ when $\lambda \geq \lambda _{cr_{1}}^{\left( 1\right) }\equiv 
\frac{2\pi (1-R)}{1+R}$. We have demonstrated in Section $3$ that
eigenvalues of the characteristic equation fulfill inequalities $0\leq
\Lambda _{1}<1$ and$\ 1<\Lambda _{2}$ $\left( \lambda >\lambda
_{cr_{1}}^{\left( 1\right) }\right) $. Therefore the fixed point $%
P_{u}^{\left( 1\right) }$ is a saddle: it has one-dimensional unstable
manifold $W_{u}\left( P_{u}^{\left( 1\right) }\right) \ $as well as
one-dimensional stable manifold $W_{s}\left( P_{u}^{\left( 1\right) }\right)
\ $in the phase space $\left( X,\ V\right) $. The basin boundary is exactly
the manifold $W_{s}\left( P_{u}^{\left( 1\right) }\right) \ $\cite{Ott1993}.

We can see in Fig. 3 fractalization of the basins for large velocities. This
phenomenon is enhanced for larger values of $\lambda $, see Fig. $4$
below.\medskip

\begin{figure}[th!]
\center \includegraphics[width=10cm, height=8cm]{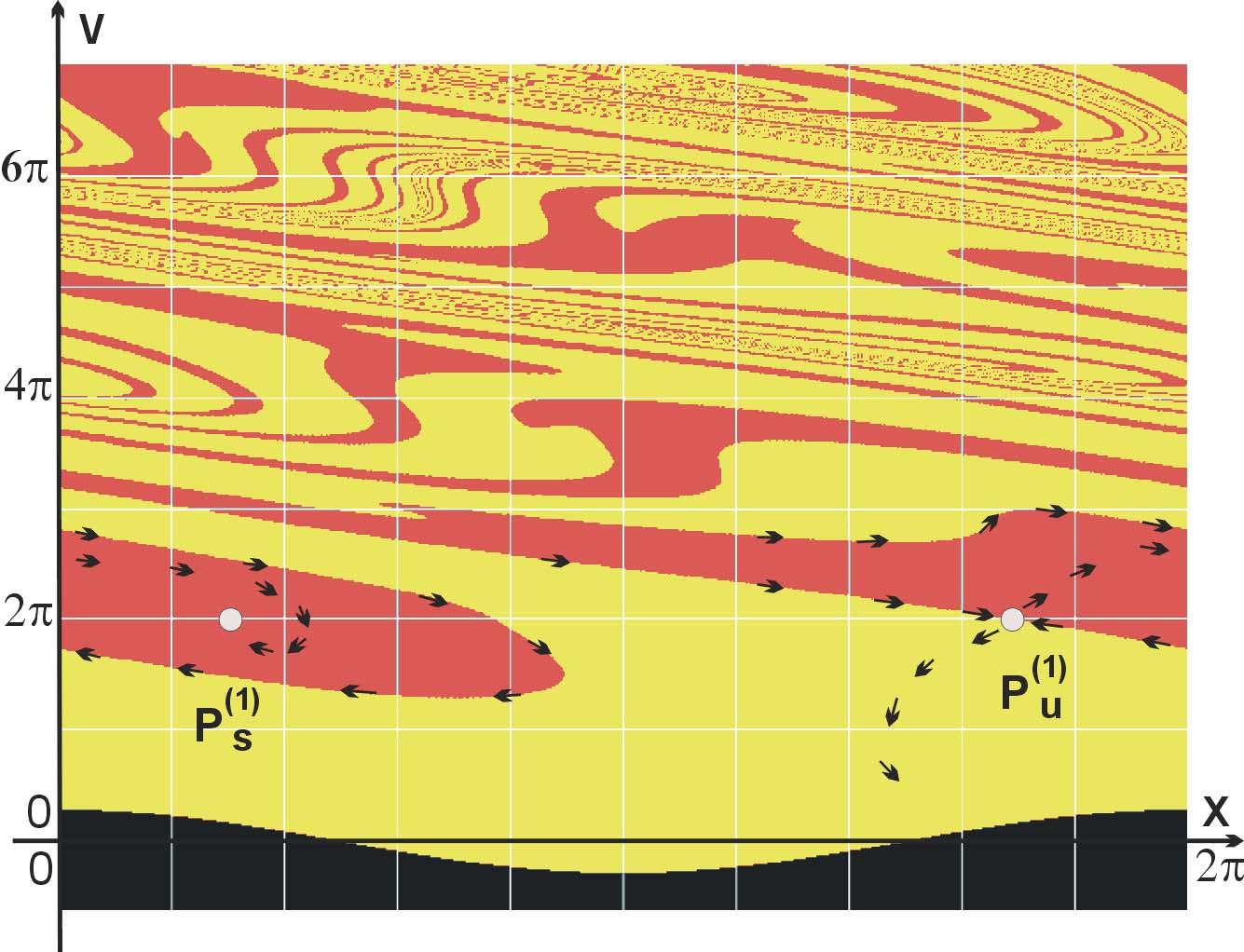}
\caption{Basins of attraction, $R=0.85,\ \lambda =0.9$.}
\label{F4}
\end{figure}

For higher values of $\lambda $, $\lambda \geq \lambda _{cr_{1}}^{\left(
2\right) }\equiv \tfrac{4\pi (1-R)}{1+R}$, the second pair of $1$ - cycles, $%
P_{s}^{\left( 2\right) }$ and $P_{u}^{\left( 2\right) }$, exist (the first
attractive and the second one a saddle). Basin boundaries are formed by $%
W_{s}\left( P_{u}^{\left( 1\right) }\right) $ and $W_{s}\left( P_{u}^{\left(
2\right) }\right) $, see Fig. $5$.\medskip

\begin{figure}[th!]
\center \includegraphics[width=10cm, height=8cm]{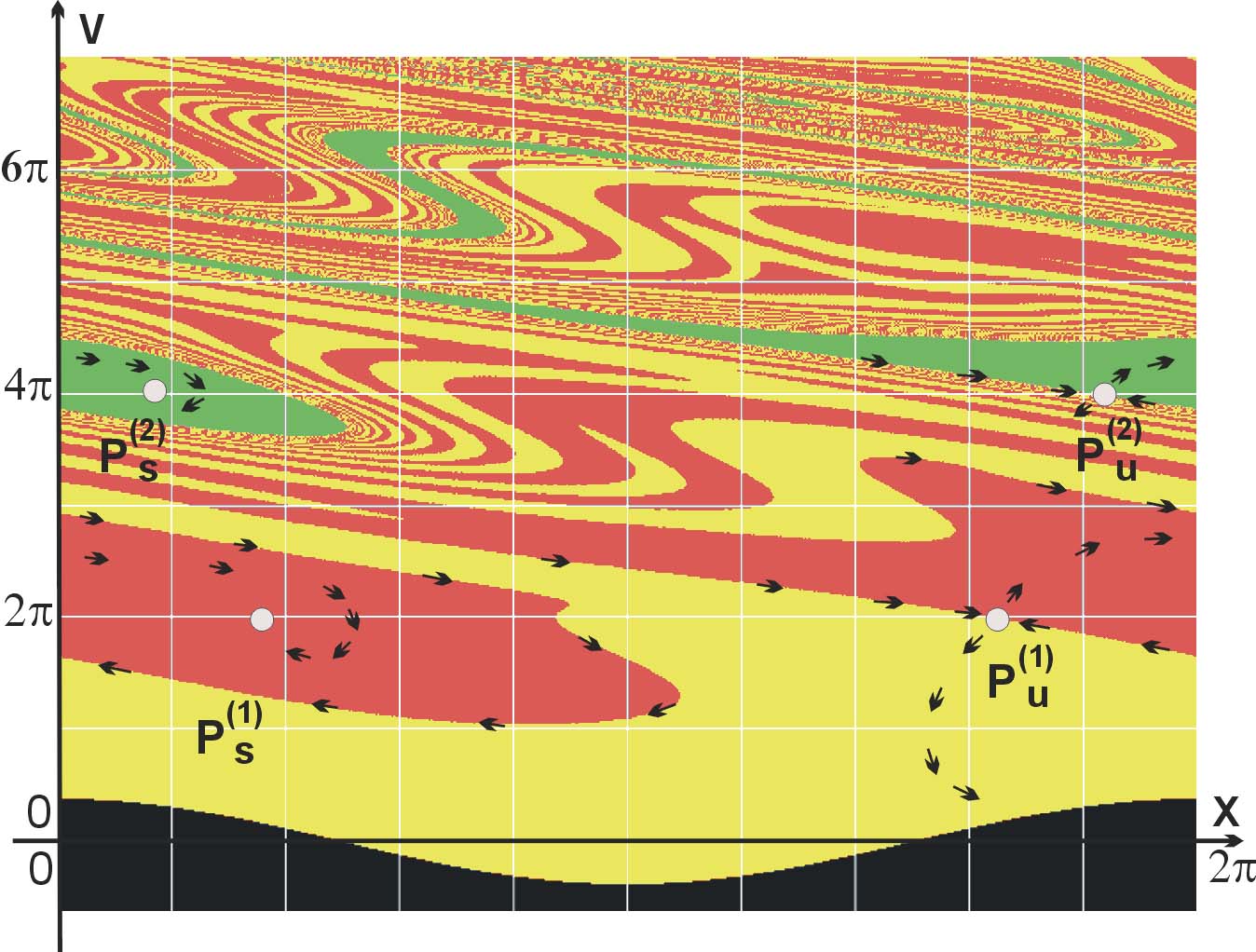}
\caption{Basins of attraction, $R=0.85,\ \lambda =1.2$.}
\label{F5}
\end{figure}

In Figs. $3,\ 4,\ 5$ fractalization of the basins of attraction is clearly
seen. In other words, the manifolds $W_{s}\left( P_{u}^{\left( 1\right)
}\right) $, $W_{s}\left( P_{u}^{\left( 2\right) }\right) $, which contribute
to the basins boundaries, become fragmented or shredded. The shredding
process of the manifolds was predicted and qualitatively described by
Whiston \cite{Wh1992}. It also seems that all three basins in Fig. $5$ are
intermingled in some parts of the phase space.

\section{Discussion}

We have analyzed dynamics of the bouncing ball via the Poincar\'{e} map. We
have found approximate solution, (\ref{maptA}), of Eq.(\ref{mapX}) defining
time of the next impact as the implicit function, $X_{i+1}=f\left( X_{i},\
V_{i}\right) $, near the grazing manifold, except from some critical points,
discussed below. It was possible to determine approximate form of the
implicit function because the series expansion of (\ref{mapX}) converges
near the grazing manifold, moreover the trivial solution, $X_{i+1}=X_{i}$,
was recognized as an initial condition and was eliminated, accordingly. The
solution (\ref{maptA}) or (\ref{maptB}) was shown be valid for more general
impacting systems, see Eq.(\ref{time}). The critical points on the grazing
manifold turn out to be zeroes of the post-impact acceleration. Let us note
here that importance of the acceleration function was stressed by Whiston
and Nordmark \cite{Wh1987, Wh1992, No1991}.

We have also studied dependence on initial conditions for the map (\ref%
{MAPXV}) in the case of coexisting attractors, namely grazing dynamics and
the first fixed point $P_{s}^{\left( 1\right) }$, as well as the second
fixed point $P_{s}^{\left( 2\right) }$. It is interesting that for initial
conditions with large velocities fractalization of basins boundaries can be
observed. This effect was predicted by Whiston \cite{Wh1992}.

There are several possible directions for further investigations. Firstly,
the role of critical points on the grazing manifold should be studied.
Secondly, the shredding process, leading to fragmentation of basins of
attraction, should be understood in detail. Finally, more general case of
impacts can be considered. Namely, it is well known from experiments as well
as from theoretical investigations that the coefficient of restitution
depends on the impact velocity, $R=r\left( \dot{x}\right) $ \cite%
{Hodkinson1835, Goldsmith1960}. There were several experimental as well as
theoretical attempts to determine dependence of the coefficient of
restitution on impact velocity, see for example \cite{Brach1991, Stronge2000}
. It is thus possible to use the approach described in the present work in
the case of velocity dependent coefficient of restitution, substituting in
the impact equation (\ref{impact2}) the constant $R$ by a function $r\left( 
\dot{x}\right) $. Alternatively, continuous model of low-speed collisions
can be considered \cite{Ivanov1996}.

\end{document}